\documentclass{SCGE}
\usepackage{newtxtext,newtxmath}
\usepackage{gensymb}
\usepackage[T1]{fontenc}
\usepackage{ae,aecompl}
\usepackage{xcolor,float,graphicx}	
\usepackage{mathrsfs}
\usepackage{breqn}
\usepackage{amsmath}	
\usepackage{amssymb}	
\usepackage{subfigure}

\newcommand{\dd}{{\rm d}}
\usepackage[colorlinks,linkcolor=blue,citecolor=blue,urlcolor=blue]{hyperref}
\let\citedash\relax
\makeatletter \providecommand{\citedash}{\hbox{-}\penalty\@m}
\makeatother
\newenvironment{seqnarray}{\small\begin{eqnarray}}{\end{eqnarray}}

\begin{document}

\begin{picture}(0,0){\rm
\put(0,-20){\makebox[160truemm][l]{\bf {\sanhao\raisebox{2pt}{.}}
Article  {\sanhao\raisebox{1.5pt}{.}}}}}
\put(0,-34){\jiuwuhao {\textcolor[rgb]{0.5,0.5,0.5}{\sf
}}}
\end{picture}

\def\bm{\boldsymbol}

\def\dl{\displaystyle}
\def\du{\end{document}}
\def\d{{\rm d}}
\def\e{{\rm e}}
\def\i{{\rm i}}

\Year{2019} 
\Month{May} 
\Vol{62} 
\No{5} 
\BeginPage{959506} 
\AuthorMark{{\rm Zhang}, et al.}  
\DOI{https://doi.org/10.1007/s11433-019-9383-y} 
\ArtNo{000000}

\title{Status and perspectives of the CRAFTS extra-galactic HI survey}

\author[1,2,3]{Kai Zhang}{zk3kw2n@nao.cas.cn}

\author[1,2,3]{Jingwen Wu}{jingwen@nao.cas.cn}
\author[1,2,3]{Di Li}{dili@nao.cas.cn}
\author[1,3]{Marko Kr\v{c}o}{}
\author[4,5]{Lister Staveley-Smith}{}
\author[1,3]{Ningyu Tang}{}
\author[1,3]{\\Lei Qian}{}
\author[1,2,3]{Mengting Liu}{}
\author[1,3]{Chengjin Jin}{}
\author[1,3]{Youling Yue}{}
\author[1,3]{Yan Zhu}{}
\author[1,3]{Hongfei Liu}{}
\author[1,3]{Dongjun Yu}{}
\author[1,3]{\\Jinghai Sun}{}
\author[1,3]{Gaofeng Pan}{}
\author[1,3]{Hui Li}{}
\author[1,3]{Hengqian Gan}{}
\author[1,3]{Rui Yao}{}
\author[]{FAST Collaboration}{}

\address[{\rm1}]{National Astronomical Observatories, Chinese Academy of Sicences, Datun Road, Chaoyang District, Beijing 10010, People's Republic of China;}
\address[{\rm2}]{University of Chinese Academy of Sciences, Beijing 100049, People's Republic of China;}
\address[{\rm3}]{CAS Key Laboratory of FAST, NAOC, Chinese Academy of Sciences, Beijing 200101, People's Republic of China;}

\address[{\rm4}]{ICRAR, The University of Western Australia, 35 Stirling Highway, Crawley,
WA 6009, Australia;}

\address[{\rm5}]{ARC Centre of Excellence for All Sky Astrophysics in 3 Dimensions (ASTRO 3D), Australia}

\maketitle \vspace{-3.5mm}{\footnotesize\begin{center} Received January 28, 2019; accepted March 4, 2019; published online March 29, 2019.
\end{center}}\vspace*{-5mm}

\begin{center}
\rule{16.5cm}{0.4pt}
\parbox{16.5cm}
{\begin{abstract} The Five-hundred-meter Aperture Spherical radio Telescope(FAST) is expected to complete its commissioning in 2019. FAST will soon begin the Commensal Radio Astronomy FasT Survey(CRAFTS), a novel and unprecedented commensal drift scan survey of the entire sky visible from FAST. The goal of CRAFTS is to cover more than 20000 $deg^{2}$ and reach redshift up to about 0.35. We provide empirical measurements of the beam size and sensitivity of FAST across the 1.05 to 1.45 GHz frequency range of the FAST L-band Array of 19-beams(FLAN). Using a simulated HI-galaxy catalogue based on the HI Mass Function(HIMF), we estimate the number of galaxies that CRAFTS may detect. At redshifts below 0.35, over $6\, \times \, 10^{5}$ HI galaxies may be detected. Below the redshift of 0.07, the CRAFTS HIMF will be complete above a mass threshold of $10^{9.5}\,M_{\odot}$. FAST will be able to investigate the environmental and redshift dependence of the HIMF to an unprecedented depth, shedding light onto the missing baryon and missing satellite problems.

\end{abstract}}
\end{center}\vspace*{-0.6cm}

\begin{center}
\parbox{16.5cm}
{\bf\jiuhao HI regions and 21-cm lines, Radio telescopes and instrumentation , Statistical and correlative studies of properties}
\end{center}

\begin{center}
{\PACS{\rm 47.55.nb, 47.20.Ky, 47.11.Fg}}
\end{center}

\textwidth=178truemm \textheight=236truemm

\wuhao\vspace*{1.5mm}

\begin{multicols}{2}

\renewcommand{\baselinestretch}{1.08} \baselineskip 12.2pt\parindent=10.8pt

\renewcommand{\thefootnote}

\section{Introduction}\label{sec:1}

HI, or neutral hydrogen is the most abundant species in the universe. Observations of the HI 21cm line can reveal a variety of information on galaxies, including their redshift, HI mass, central velocity, dynamical mass, and so on ~\cite{Roberts1975}. Wide-field blind extragalactic HI sky surveys can be implemented owing to the advent of multi-beam receivers which greatly improve the efficiency and allow the surveys to cover cosmologically significant volumes. From the HIPASS(HI Parkes All Sky Survey~\cite{Barnes2001}~\cite{Meyer2004}), to the recently completed ALFALFA (Arecibo Legacy Fast ALFA~\cite{Giovanelli2005}~\cite{Haynes2011}~\cite{Giovanelli2015}~\cite{Haynes2018}) survey, blind surveys have played a key role in exploring the HI distribution and properties of nearby galaxies.

The HI mass function (HIMF) depicts the cosmic number density per bin of HI mass~\cite{Briggs1990} and is found to be well fitted by the Schechter function~\cite{Schechter1976}. Faint-end slopes of the HIMF are closely related to the
"missing satellite problem"~\cite{Klypin1999}~\cite{Moore1999}. At high redshift (z>1.5) the total mass density of HI, or $\Omega_{\rm HI}$, can be deduced from the absorption lines of quasar optical spectra through damped-Lyalpha objects~\cite{Storrie2000}~\cite{Peroux2003}, while at lower redshifts 
$\Omega_{\rm HI}$ can be evaluated by the  HIMF~\cite{Zwaan2005}~\cite{Martin2010}~\cite{Jones2018}. The HIMF is very important in understanding the galactic HI evolution as a function of redshift, and serves as a test of theoretical cosmological simulations~\cite{Obreschkow2009}~\cite{Duffy2012a}. With large sampled populations and spatial volumes blind HI surveys are ideal for measuring the HIMF and for studying the environmental dependence of the  HIMF~\cite{Zwaan2005}~\cite{Moorman2014}~\cite{Jones2016b}~\cite{Jones2018}. FAST and ASKAP (The Australian Square Kilometre Array Pathfinder) will carry out the next generation of HI blind sky surveys~\cite{Duffy2008}~\cite{Duffy2012}, to enable the study of the evolution of the HIMF and $\Omega_{\rm HI}$ up to higher redshifts.

As the largest filled-aperture single-dish telescope in the world, FAST~\cite{Nan2006}~\cite{Nan2011} is expected to finish its commissioning in 2019~\cite{Li2018}~\cite{Jiang2019}. The Commensal Radio Astronomy FasT Survey (CRAFTS)~\cite{Li2018} will simultaneously conduct an extra-galactic HI, Galactic HI imaging, pulsar search, and Fast Radio Burst (FRB) search surveys. The predictions~\cite{Duffy2008} show FAST would be an excellent instrument for large-scale HI surveys. Based on test observations taken during FAST commissioning, we use updated parameters of FAST and the latest HIMF from ALFALFA survey to predict the capability of CRAFTS to detect HI galaxies.

We briefly introduce the CRAFTS survey plan in \ref{sec:2}, and summarize several key parameters of FAST in \ref{sec:3}. The sensitivity of CRAFTS with regards to the extragalactic HI survey is discussed in \ref{sec:4}. The number of HI galaxies CRAFTS may detect is discussed in \ref{sec:5}. We will discuss the impact of confusion on CRAFTS in \ref{sec:6}, and a summary is provided in \ref{sec:7}. The cosmological distances calculation and cosmological corrections in this paper refer to ~\cite{Hogg1999} and ~\cite{Meyer2017}, and we assume $\rm H_{0}(Hubble\, Constant)=73 \, \rm km\,\rm s^{-1} \, \rm Mpc^{-1}$, $\rm \Omega_{m}(Density\, parameter\, of \,matter)=0.25$ and $\rm \Omega_{\Lambda}(Density\, parameter\, of\, dark\, energy)=0.75$.

\section{The CRAFT Survey Plan}\label{sec:2}
When inactive, FAST's surface is a partial sphere composed of over 4400 panels connected to over 2250 actuators with an aperture of 500m.  While observing, the actuators behind the reflector deform the illuminated surface into a 300m-aperture paraboloid, with an rms(root mean square) uncertainty of $\sim$4mm which is sufficient for observations at frequencies up to 5GHz\cite{Nan2006}. In the case of FAST, the most efficient way to conduct a large commensal survey is to use drift scans which have the added benefit of minimizing gain fluctuations due to variations in the surface and feed position. CRAFTS will observe galactic HI, extra-galactic HI, pulsars, and FRBs simultaneously while drifting. CRAFTS will conduct two full passes surveys similarly to ALFALFA~\cite{Giovanelli2005}, which should help alleviate the influence of Radio Frequency Interference (RFI) and make low SNR (Signal-to-Noise Ratio) detections more reliable. It will take CRAFTS approximately 220 full days of observation to complete each pass~\cite{Li2018}.

\begin{figure}[H]

	\hspace*{-0.7cm}\includegraphics[scale=0.45]{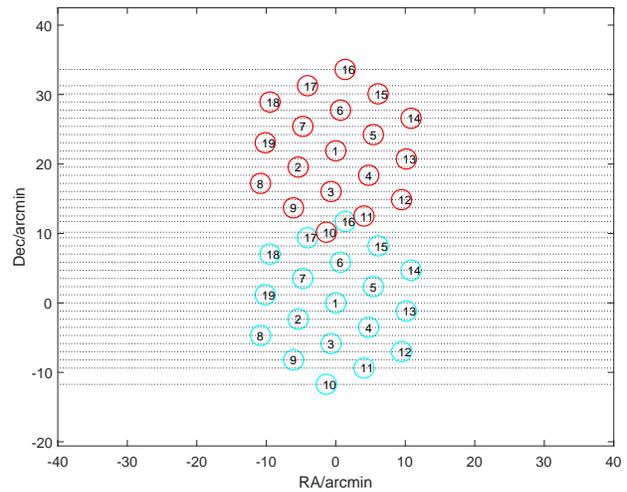}
    \caption{An example sketch of two adjacent drift scans of FLAN using the orientation used by CRAFTS. Blue and red circles with a diameter of 2.9 arcmin represent the position and the size of beams in two drift scans. The dotted lines show the drifting tracks of individual beams.}
    \label{fig:scan pattern}
\end{figure}

While drifting, the 19-beam receiver (the FAST L-band Array of 19-beam, or FLAN), will be rotated by $ 23.4\degree$ to achieve a super-Nyquist sampling. The spacing of each scan would be $ 21.9'$  in declination to fill the gap of two outer beams so that the survey would be nearly uniformly covered as described in \cite{Li2018}. Figure~\ref{fig:scan pattern} shows a sketch of two adjacent drift scans of FLAN using this orientation.

\section{Key FAST Parameters}\label{sec:3}
FAST is located at a latitude of $ 25.6529 \degree \rm N$. The effective illuminated aperture size is about 300 m up to a zenith angle of $ 26.4 \degree $, and can be partially illuminated up to FAST's maximum zenith angle of $40 \degree$. Thus the total sky visible from FAST is approximately $20000 \ \rm deg^{2}$  within a declination range between $-14\degree$ and $66 \degree$.

The FLAN bandwidth (1.05 GHz to 1.45 GHz) corresponds to a maximum redshift of 0.35 for the 21 cm HI line. The channel width of FLAN for the extragalactic HI survey is about 7.6 kHz, corresponding to a velocity resolution of 1.6 km/s for HI.

We use FAST commissioning data to estimate its beamsize, gain and system temperature. Much of the data are taken from ~\cite{Jiang2019} and the parameters are all obtained from center beam of the FLAN.

\begin{figure*}
	\hspace*{0cm}\vspace{-0cm}\includegraphics[scale=0.5]{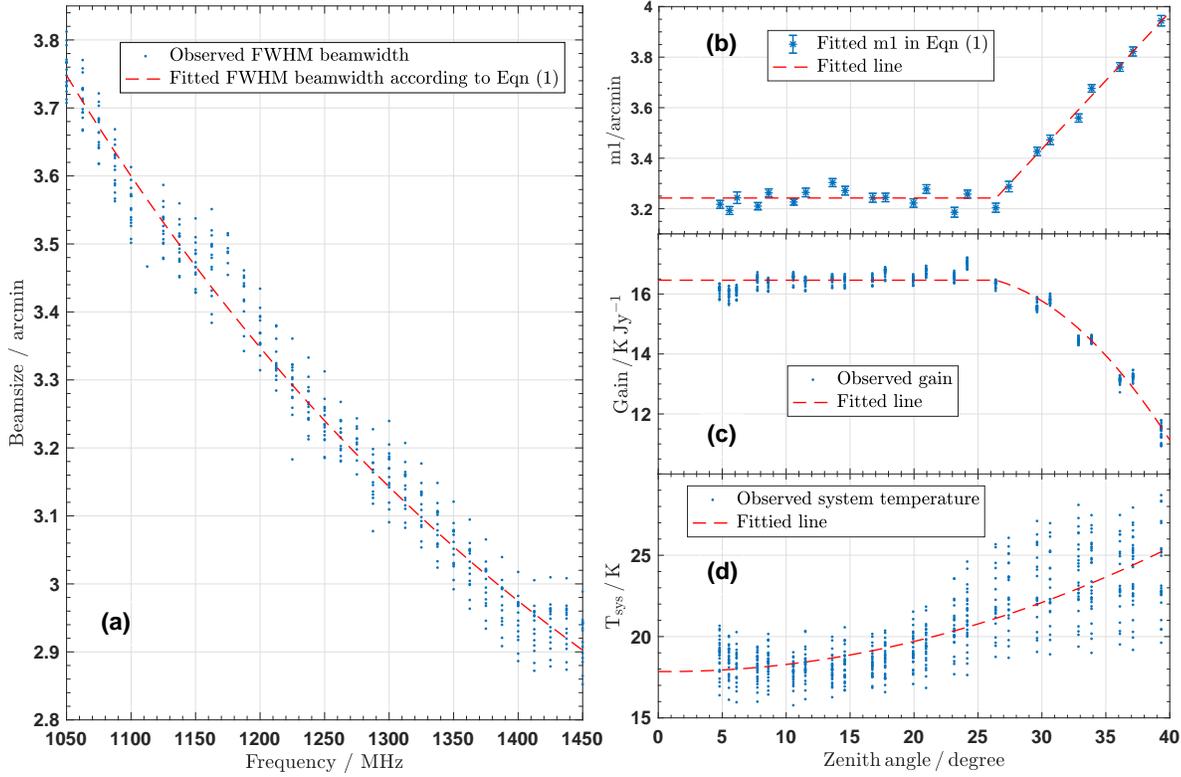}
    \caption{Panel (a): beamsize versus frequency when zenith angle is below $26.4 \degree$, the red dashed line represents the fitted result from equation~(\ref{eq:theta1}). Panel (b): the blue asterisks represent the fitted result of m1 by equation~(\ref{eq:theta1}) at different zenith angles, which tells the beamsize at the frequency of 1250 MHz, while the red dashed line represents the fitted line from equation~(\ref{eq:theta2}).  Panel (c): the gain versus zenith angle, the red dashed line represents the fitted result from equation~(\ref{eq:gain}). Panel (d):system temperature versus zenith angle,the red dashed line represents the fitted result from equation~(\ref{eq:Tsys}). The parameters here are all obtained from center beam of the 19-beam receiver(FLAN).}
    \label{fig:parameter}
\end{figure*}

Due to the wide bandwidth of FLAN and the limited feed cross-section size, the electric field of the aperture antenna tends to be uniformly distributed at low frequencies. While at high frequencies the distribution function of electric field that illuminates the aperture resembles a Gaussian function. Therefore the FLAN beam size cannot be simply estimated by assuming it is inversely proportional to frequency; at low frequencies the beamsize is smaller than otherwise predicted. The FLAN FWHM beamsize of center beam can be accurately estimated by
\begin{seqnarray}
\label{eq:theta1}
\left(\dfrac{\theta}{\rm '}\right)=\rm \left[ \frac{m1(ZA)}{'}\right]\left(\dfrac{\dd Freq/1250}{\rm MHz}\right)^{-1}+\rm \left(\frac{m2}{'}\right)\left[\left(\dfrac{\rm Freq}{\rm MHz}\right)-1250\right],
\end{seqnarray}

where $\theta$ is beamsize in arcmin, Freq is frequency in MHz, m1 is beamsize at the frequency of 1250 MHz in arcmin and m2 is the correction factor in arcmin. From observations, m1 is approximately 3.24 arcmin and m2 is approximately 5.475$\times 10^{-4}$ arcmin when zenith angle is below $ 26.4 \degree $. Panel (a) in Fig.~\ref{fig:parameter} shows how beamsize varies with frequency when zenith angle is below $ 26.4 \degree $.

For zenith angles above $ 26.4 \degree $ the aperture is only partially illuminated and thus the beamsize will increase. When zenith angle increases, the illuminated aperture size will decrease, we assume that the illuminated aperture size decreases linearly when zenith angle increases and thus m1 is proportional to zenith angle. For m2 we assume it doesn't change. Panel(b) in Fig.~\ref{fig:parameter} shows the relation between m1 and zenith angle. Thus when zenith angle is above $ 26.4 \degree $, m1 can be approximated by
\begin{seqnarray}
\label{eq:theta2}
{\rm\left( \frac{m1}{'}\right)}=3.24+5.41 \times 10^{-2}\left[\left(\dfrac{\rm ZA}{\,\degree}\right)-26.4\right],
\end{seqnarray}
where ZA is zenith angle in degree.

The gain is about 16.46 K/Jy for the center beam when zenith angle is below $26.4 \degree$ and is largely independent of frequency over the bandwidth covered by FLAN. Gain decreases for zenith angles greater than $26.4 \degree$ due to the surface being only partially illuminated as shown in Panel (c) in Fig.~\ref{fig:parameter}. When zenith angle is above $26.4 \degree$, the gain may be estimated by a second degree polynomial such as 

\begin{seqnarray}
\label{eq:gain}
\left(\frac{\rm G}{\rm K/Jy}\right)=16.46-0.02\left[\left(\frac{\rm ZA}{\,\degree}\right)-26.4\right]^{2}-0.12\left[\left(\frac{\rm ZA}{\,\degree}\right)-26.4\right],
\end{seqnarray}
where G is the gain in K/Jy and ZA is zenith angle in degree. At a zenith angle of $40 \degree$, the gain will decrease to about 11.0 K/Jy, with a loss of  approximately $33\%$.

Panel (d) in Fig.~\ref{fig:parameter} shows the relation between system temperature and zenith angle. We use a second-order polynomial function to estimate the relation between system temperature and zenith angle with 
\begin{seqnarray}
\label{eq:Tsys}
\left(\frac{T_{\rm sys}}{\rm K}\right)=4.9 \times 10^{-3} \left(\frac{\rm ZA}{\,\degree}\right)^{2}+5.5 \times 10^{-3}\left(\frac{\rm ZA}{\,\degree}\right)+17.85,
\end{seqnarray}
where $T_{\rm sys}$ is system temperature in K, and ZA is zenith angle in degree. At zenith angles above $26.4 \degree$ the feed is tilted back slightly so as to avoid spill over. However, radiation from the surrounding mountain peaks enters the nearer sidelobes and increases the system temperature. A ground screen made from wire mesh was proposed to reduce $\rm T_{sys}$ for FAST when the zenith angle is large, which is designated as the $\textquoteleft$backward illumination\textquoteright \, mode~\cite{Jin2013}.

\begin{figure}[H]

	 \hspace*{-0cm}\vspace{-0cm}\includegraphics[scale=0.57]{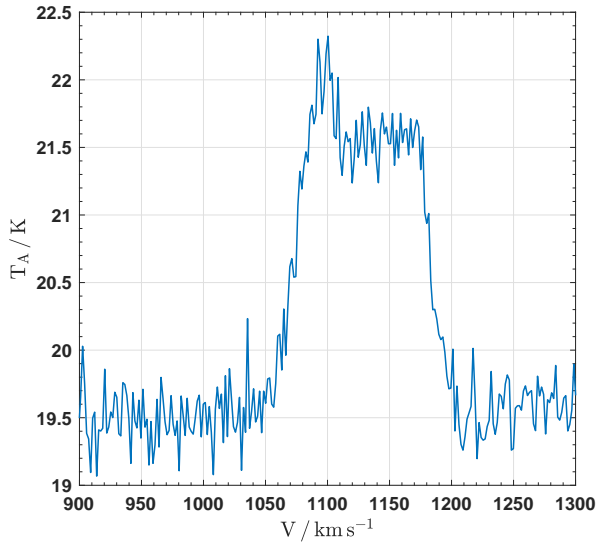}
    \caption{A spectrum of AGC11820 taken by FAST on 2018/10/22. The horizontal axis represents radial velocity and the vertical axis represents antenna temperature.}
    \label{fig:AGC11820}
\end{figure}

Fig.~\ref{fig:AGC11820} shows an example spectrum of AGC11820 taken by FAST on 2018/10/22.

\section{Sensitivity Calculation}\label{sec:4}

\subsection{Effective Integration Time}

The effective integration time of a drift scan survey reflects the equivalent rms noise by combining data from multiple observations, which may be estimated by
\begin{seqnarray}
\label{eq:teff}
t_{\rm obs}/t_{\rm eff}=\Omega_{\rm obs}/ \Omega_{\rm b},
\end{seqnarray}
where $t_{\rm eff}$ is the effective integration time for each point in the sky within the observed region, $t_{\rm obs}$ is the total observation time, $\Omega_{\rm obs}$ is the total solid angle of observation, and $\Omega_{\rm b}$ is the solid angle of the beam. 
\begin{figure}[H]

	 \hspace*{0cm}\vspace{0cm}\includegraphics[scale=0.51]{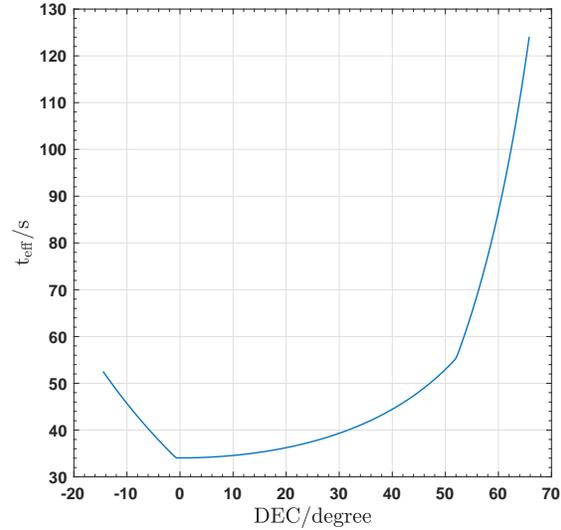}
	 \hspace*{0cm}\vspace{-0cm}\includegraphics[scale=0.51]{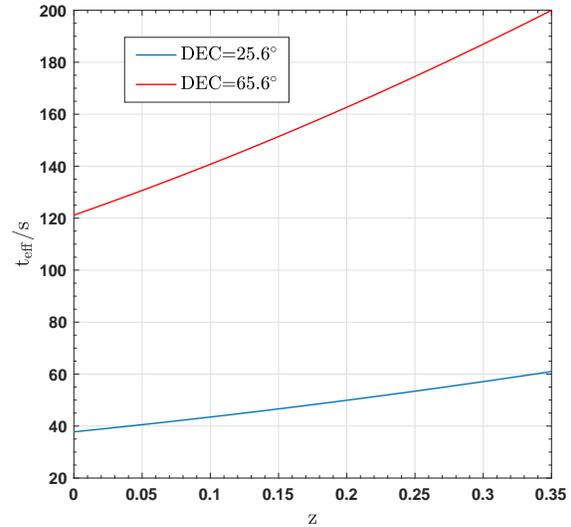}
    \caption{The upper figure shows $t_{\rm eff}$ versus DEC at redshift of 0 after one pass. The bottom figure shows $t_{\rm eff}$ after one pass as a function of z at DEC of $25.6 \degree$ and $65.6 \degree$ in blue and red solid line respectively.}
    \label{fig:teff}
\end{figure}

We assume all 19 beams of FLAN have same beam pattern, which can be estimated as an axisymmetric Gaussian function, so $\Omega_{\rm b}=19 \times 1.13 \times \theta^{2}$, where $\theta$ is the beamsize. If we assume the drifting speed at certain declination is constant and at the equator the drifting speed reaches its maximum, which is roughly 0.25 arcmin/s, thus the drifting speed of survey $v_{\rm s}/\rm arcmin\ \rm s^{-1}=0.25*cos(DEC) $, where DEC is the declination of the center beam of FLAN while drifting. $\Omega_{\rm obs}$ can be represented by $\delta v_{\rm s} t_{\rm obs}$, where $\delta = 21.9' $ is the spacing in declination between two adjacent drift scans. Thus $t_{\rm eff}$ can be estimated by
\begin{seqnarray}
\label{eq:teff}
\left(\frac{t_{\rm eff}}{s} \right)=\frac{19 \times 1.13 \times \left(\frac{\theta}{'} \right) ^{2}}{  \left(\frac{0.25}{'/s}\right) \times {\rm cos} { \left(\frac{\rm DEC}{\degree}\right)  }  \left(\frac{\delta}{'}\right)  }.
\end{seqnarray}

If we overlook the peculiar motion of the earth and HI source, the frequency of HI line we observed is: $\nu_{\rm obs}=\nu_{\rm HI}/(1+{\rm z})$, where $\nu_{\rm HI} = 1420.4 \rm MHz$ is the rest-frame HI frequency, and $\rm z$ is the redshift of the source. The upper plot in Fig.~\ref{fig:teff} shows the effective integration time after one pass as a function of DEC at redshift of 0. The main factors that influence $t_{\rm eff}$ are the beamsize and $1/{\rm cos}({\rm DEC})$. When DEC ranges between  $-1.2 \degree$ and $52 \degree$, $\theta$ doesn't vary with DEC, so $t_{\rm eff}$ will change with $1/{\rm cos}({\rm DEC})$. When DEC is below $-1.2 \degree$ or above $52 \degree$, the beamsize would increase linearly with ZA, so the growth rate of $t_{\rm eff}$ will become higher at higher zenith angle. The bottom plot of Fig.~\ref{fig:teff} shows how $t_{\rm eff}$ varies with z at two typical DECs after one pass.

\begin{figure}[H]

	\hspace*{-0cm}\vspace{-0cm}\includegraphics[scale=0.5]{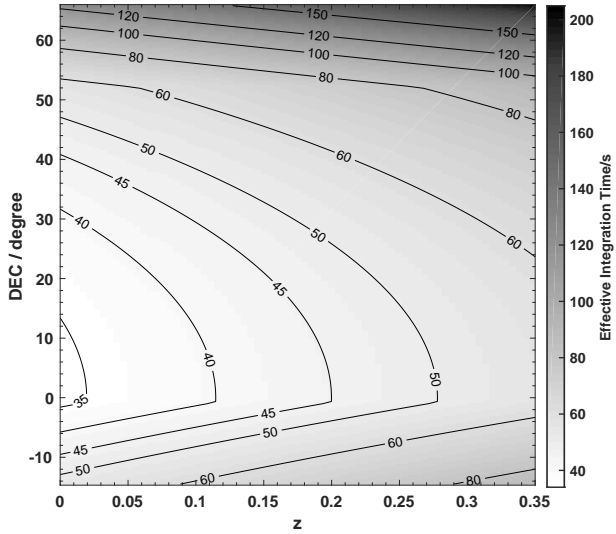}
    \caption{The effective integration time map of one-pass of CRAFTS in z-DEC plane. The pixel size is $21.9'$ in declination and 0.005 in redshift.}
    \label{fig:teff_map}
\end{figure}

Fig.~\ref{fig:teff_map} shows the effective integration time map of one-pass of CRAFTS in z-DEC plane based on equation~(\ref{eq:teff}). The map shows that in most CRAFTS regions, $t_{\rm eff}$ is from 35s to 60s, while at high zenith angle, $t_{\rm eff}$ will increase rapidly, especially at high declination regions, where the drifting speed is slower thus the duration time of source in beam will increase. CRAFTS plans to carry out a two-pass survey, so the ultimate effective integration time is twice the value in Fig.~\ref{fig:teff_map}.

\begin{figure}[H]

	 \hspace*{0cm}\vspace{0cm}\includegraphics[scale=0.53]{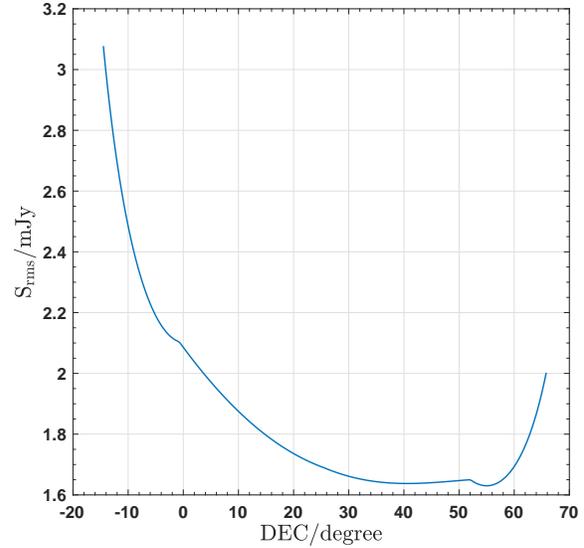}
	 \hspace*{0cm}\vspace{-0cm}\includegraphics[scale=0.53]{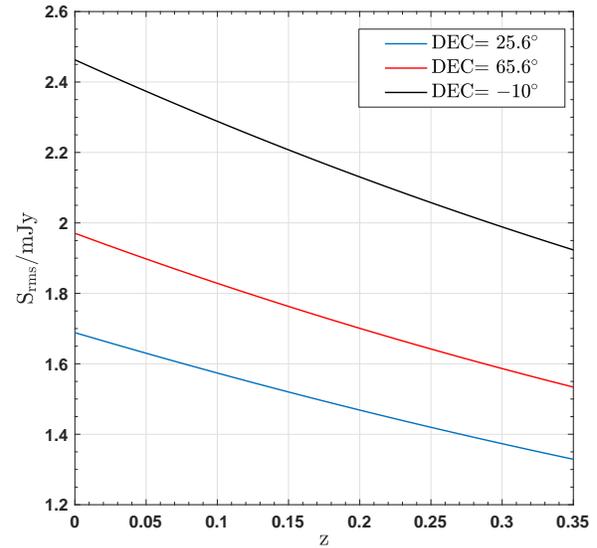}
    \caption{The upper figure shows $S_{\rm rms}$ as a function of DEC at redshift of 0 after one pass. Bottom figure shows $S_{\rm rms}$ after one pass versus z at DEC of $25.6 \degree$, $65.6 \degree$, $-10 \degree$ in blue, red and black soild line respectively.}
    \label{fig:Srms_map}
\end{figure}

\subsection{Sensitivity per Channel}

The rms noise per channel or $S_{\rm rms}$, can be estimated by~\cite{Giovanelli2015}
\begin{seqnarray}
\label{eq:sensitivity}
S_{\rm rms}=\frac{T_{\rm sys}}{\rm G}\frac {1}{\sqrt{2f_{\rm t}\Delta f_{\rm ch}t_{\rm s}}},
\end{seqnarray}

where $\rm T_{sys}$ is system temperature in K, $\rm G$ is gain of the telescope in $\rm K/Jy$, $\Delta f_{\rm ch}$ is single channel width in Hz and $t_{\rm s}$ is integration time in s, here we can use the effective integration time we calculated before. The factor $f_{\rm t}$ accounts for observation details like bandpass subtraction method, spectral smoothing and so on. For ALFALFA, $f_{\rm t} \approx 0.7$, here we adopt this value for CRAFTS extragalactic HI survey.

The top figure in Fig.~\ref{fig:Srms_map} shows $S_{\rm rms}$ as a function of DEC at redshift of 0. The main factor that influences $S_{\rm rms}$ at ZA below $25.6 \degree$ is the effective integration time: $S_{\rm rms}$ will decrease along the increase of DEC. However, at high ZA, the increase of gain and $T_{\rm sys}$ will cause $S_{\rm rms}$ increase at a relatively higher rate. The bottom figure in Fig.~\ref{fig:Srms_map} shows $S_{\rm rms}$ varies as a function of z at three typical DECs.

\begin{figure}[H]

	 \hspace*{-0cm}\vspace{-0cm}\includegraphics[scale=0.5]{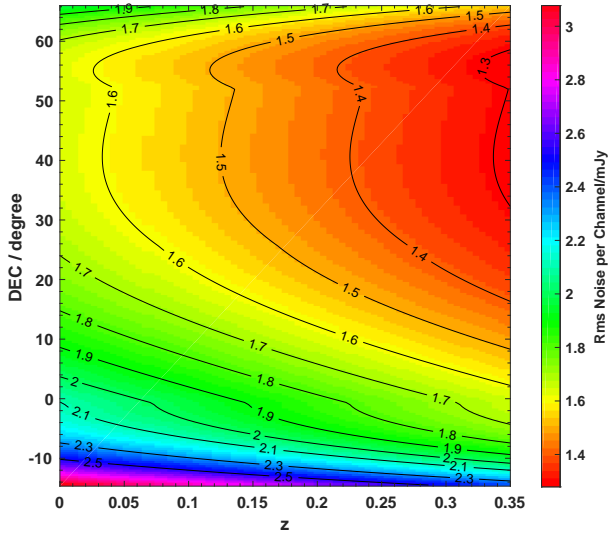}
    \caption{The rms noise per channel map for CRAFTS drifting one-pass in z-DEC plane, with same pixel size as Fig.~\ref{fig:teff_map}. }
    \label{fig:Srms_map1}
\end{figure}

$S_{\rm rms}$ of one-pass of CRAFTS in z-DEC plane is shown in Fig.~\ref{fig:Srms_map1}. Fig.~\ref{fig:Srms_map1} implies that $S_{\rm rms}$ will not fluctuate rapidly with DEC when DEC is above $0 \degree$. At higher redshift, the difference of $S_{\rm rms}$ between two adjacent drift scans will become larger. At DEC below $10 \degree$, $S_{\rm rms}$ will increase rapidly when DEC decreases, which is caused by relatively low $t_{\rm eff}$, high $T_{\rm sys}$ and low gain. For two-pass of CRAFTS, $S_{\rm rms}$ can be obtained by dividing the value in Fig.~\ref{fig:Srms_map1} by square root of 2.

\subsection{Flux Limit for HI Detection}

The velocity line width of HI signal at source rest frame or $ \Delta V_{\rm rest} $, can be estimated by~\cite{Meyer2017}
\begin{seqnarray}
\label{eq:V_rest}
\Delta V_{\rm rest} \simeq \frac{c(1+{\rm z})}{\nu_{\rm HI}}\Delta \nu_{\rm obs},
\end{seqnarray}where $\Delta \nu_{\rm obs} $ is the frequency line width of signal we observe, c is the speed of light and z is the redshift of the source. Here we assume the peculiar motion speed and line width of the source are much less than c. The relation between line width in source\textquoteright s rest frame, $\Delta V_{\rm rest}$, and in the observed frame, $\Delta V_{\rm obs}$, can be donated by $\Delta V_{\rm rest}=\Delta V_{\rm obs}/\left(1+{\rm z}\right)$. Thus the corresponding observed velocity width in a single channel with channel width of $ \Delta f_{\rm ch}$ while observing an HI source at redshift of z, can be estimated by
\begin{seqnarray}
\label{eq: V_ch}
\Delta V_{\rm ch} \simeq \frac{c(1+{\rm z})^{2}}{\nu_{\rm HI}}{\Delta f_{\rm ch}}.
\end{seqnarray}

The SNR of a signal can be estimated by~\cite{Haynes2018}

\begin{seqnarray}
\label{eq:SNR}
SNR = \left(\frac{S^{V_{\rm obs}}}{{\rm Jy}\,{\rm km}\,{s^{-1}}}\right)\left(\frac{\Delta V_{\rm obs}}{{\rm km}\,{\rm s^{-1}}}\right)^{-1}{f^{1/2}_{\rm smo}}\left(\frac{S_{\rm rms}}{\rm Jy}\right)^{-1},
\end{seqnarray}
where $S^{V_{\rm obs}}$ is the integrated flux density of HI line in observed frame, we can make an assumption that HI line profile is a top-hat function, so $S^{V_{\rm obs}}=S_{\rm peak} \Delta V_{\rm obs}$, where $S_{\rm peak}$ is peak flux of HI line and $\Delta V_{\rm obs}$ is the line width of HI line in observed frame. Usually $\Delta V_{\rm obs}$ can be described by $W_{50}$, which is measured at $50 \%$ level of each of two peaks. $f_{\rm smo}$ is the number of independent channels that signals can be smoothed over. For very broad HI signal profiles, the flux at two horns is much larger than that at center part of profiles, so direct smoothing will give diminishing returns~\cite{Jones2015}. We adopt the form of $f_{\rm smo}$ in ~\cite{Jones2015}, which used the transition of completeness function of ALFALFA survey as the cutoff of $f_{\rm smo}$ and smoothed the signal to full line width to calculate SNR. $f_{\rm smo}$ is adopted as $\Delta V_{\rm obs}/ \Delta V_{\rm ch}$ while $\Delta V_{\rm obs} < 10^{\rm 2.5} \ \rm km \ \rm s^{-1}$ and $ 10^{2.5}/ \Delta V_{\rm ch} $ while $\Delta V_{\rm obs} \geq 10^{\rm 2.5} \ \rm km \ \rm s^{-1}$, where $\Delta V_{\rm ch}$ is the corresponding observed HI line velocity width in a single channel.

The flux limit of the telescope while observing HI galaxies at a given SNR can be denoted by

\begin{seqnarray}
\label{eq:S_lim}
S_{\rm lim} = SNR * \frac{S_{\rm rms}}{f_{\rm smo}^{1/2}}.
\end{seqnarray}
If the peak flux of an HI galaxy signal is above $S_{\rm lim}$, we can say the telescope could detect that galaxy.

~\\
~\\
~\\

\section{Estimating the HI Galaxy Count}\label{sec:5}

The HIMF tells us the relation between HI mass and number density of HI galaxy in our universe, which can be well fitted by Schechter function~\cite{Schechter1976}. The HIMF could be expressed in the form of
\begin{seqnarray}
\label{eq:HIMF}
\phi(M_{\rm HI}) &=& \frac{dN_{\rm gal}}{dV \,{ dlog}_{10}(M_{\rm HI)}) } \,, \\
	&=& {\rm ln}(10)\, \phi_{*}\, (\frac{M_{\rm HI}}{M_{*}})^{\alpha +1}\, e^{-(\frac{M_{\rm HI}}{M_{*}})}\, ,
\end{seqnarray}
where $dN_{\rm gal}$ is the average number of galaxies in comoving volume element $dV$, $\phi_{*}$ is the normalization constant, $M_{*}$ is the \textquoteleft knee mass\textquoteright \,and $(\alpha+1)$ is the low-mass slope, which is usually referred as the \textquoteleft faint end\textquoteright \,of HIMF.

The HI mass of a galaxy, $ M_{\rm HI}$, at redshift of z can be expressed by~\cite{Roberts1975}~\cite{Meyer2017}
\begin{seqnarray}
\label{eq:mass limit}
\frac{M_{\rm HI}}{M_{\odot}} = \frac{2.35 \times 10^{5}}{(1+{\rm z})^{2}} \left[ \frac{d_{\rm L}({\rm z})}{\rm Mpc} \right]^{2} \left(\frac{S^{V_{\rm obs}}}{ {\rm Jy}\, {\rm km}\,{\rm s^{-1}}} \right),
\end{seqnarray}
where $ d_{\rm L}(z) $ is the luminosity distance to the galaxy and $S^{V_{\rm obs}}$ is the velocity integral flux in observed frame, which can be approximated by $ S_{\rm peak}W_{50} $ if we assume HI signal is a top-hat function. We can substitute $S_{\rm peak}$ by $S_{\rm lim}$ in equation~(\ref{eq:S_lim}) to calculate the minimum HI mass or $M_{\rm HI,lim}$ FAST will detect. If we assume the sensitivity of ALFALFA is 2.4 mJy~\cite{Jones2018} at resolution of $10 \, {\rm km} \, \rm s^{-1}$, for detecting an HI galaxy with a line width of $200 \, {\rm km} \, \rm s^{-1}$ at redshift of 0.05, in ALFALFA survey, $M_{\rm HI,lim}$ will be about $10^{10} M_{\odot}$; while for two-pass CRAFTS, $M_{\rm HI,lim}$ will be about $10^{9.3} M_{\odot}$ and $10^{9.5} M_{\odot}$ at DEC of $25.6 \degree$ and $-14.0 \degree$, respectively.

The method we use here to make mock catalogue is similar to~\cite{Jones2015}, we integrate the HIMF ranges from 6.2 to 14 times of ${\rm log10} \,(M_{\rm HI}/M_{\odot})$ to obtain the average HI galaxy number density and use the normalized HIMF as the probability distribution function (PDF) of $M_{\rm HI}$. We make a mock catalogue by using the PDF we obtain, the number of galaxy in mock catalogue is obtained by multiplying average number density with the volume of sky region we plan to observe. CRAFTS can detect galaxies with $M_{\rm HI}$ above $M_{\rm HI,lim}$ in mock catalogue. Following~\cite{Duffy2008}, we assume the observed line width of all galaxies is $200 \, {\rm km} \, \rm s^{-1}$, which is a close approximation of average velocity width of HI galaxies.

Using this method, we derive that the total number of HI galaxies CRAFTS will detect in one-pass and two-pass is approximately $4.0\, \times \, 10^{5}$ and $6.5\, \times \, 10^{5}$, respectively. The median redshift for CRAFTS after two passes is close to 0.07, and the HI mass limit for detecting a galaxy with line width of $200 \, {\rm km} \, \rm s^{-1}$ at that redshift is approximately $10^{9.5} \, M_{\odot}$. CRAFTS may not be able to detect HI galaxies efficiently at redshift above 0.2. We may need more passes or some strategies to increase SNR to reach higher redshift for HI galaxy detection. A detailed description and discussion of our simulations and the methodology in deriving the total number of HI galaxies in CRAFT survey will be presented in a separated paper (Zhang et al. in prep.).

 For the determination of \textquoteleft faint end\textquoteright \,of HIMF, in ~\cite{LiJ2018}, the author discussed the potential of FAST in detecting low mass HI galaxies in local group, from their simulation results, two-pass CRAFTS would detect 32 galaxies with HI mass above $10^{5} M_{\odot}$ by assuming the effective integration time is 60s, future detection on low mass galaxies will be a complementary test for current simulation models.

\section{Confusion}\label{sec:6}

Confusion is caused by source blending in the beam. Beamsize and HI mass limit will increase at high redshift, which will grow the impact of confusion to HI surveys. Confusion will cause biases on flux, line width and redshift measurement of HI galaxies, which will thus influence the calculation of survey products like correlation functions(CF), HIMF and HI line width functions(WF)~\cite{Jones2015}. According to previous study~\cite{Duffy2008}~\cite{Jones2016a}, confusion might be the main factor that limits survey capability of FAST at high redshifts due to its relatively large beam size. 

Here we use the model in ~\cite{Jones2016a} to analyze the impact of confusion on CRAFTS. That work created a mock stack in the sky with velocity range of $600 \,\rm km \, s^{-1}$, which is the broadest velocity width of HI galaxy. They defined a cylinder in redshift space centered on the target being stacked, the radius of the cylinder is the length of beam size projected on the sky. Then they calculated the total HI mass in the volume of cylinder as an estimation of the confused HI mass that can not be separated from the spectrum of the target, which is referred to $\textquoteleft$confused mass\textquoteright \, or $M_{\rm conf}$. We compare this confused mass to the HI mass limit CRAFTS could detect. If $M_{\rm conf}$ is close to $M_{\rm HI,lim}$, it will be difficult for CRAFTS to tell whether there is confusion source in the detected signal or not.

\begin{figure}[H]

	 \hspace*{-0cm}\vspace{-0cm}\includegraphics[scale=0.59]{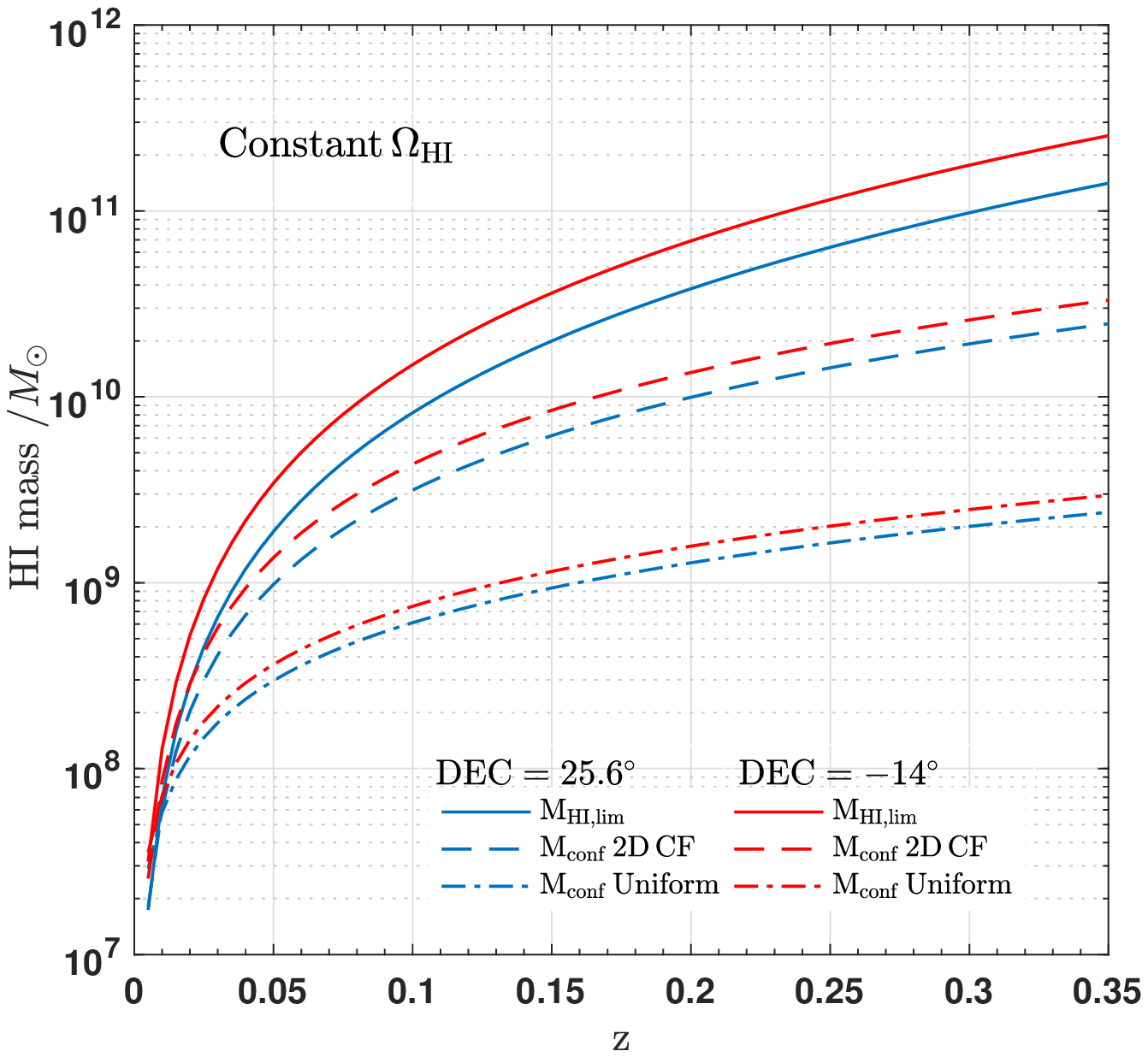}
	 \hspace*{-0cm}\vspace{-0cm}\includegraphics[scale=0.59]{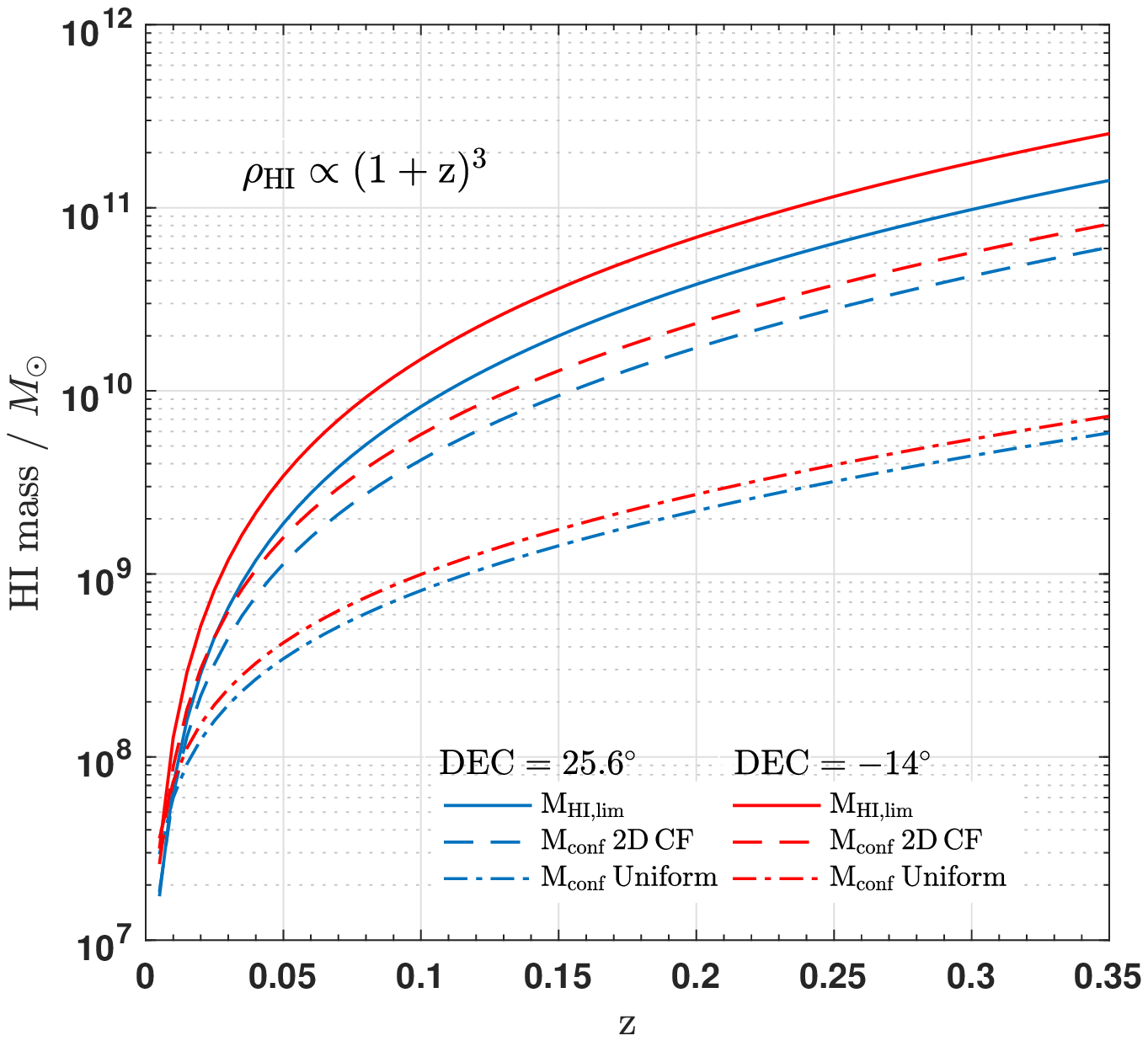}
    \caption{The comparison of $M_{\rm HI,lim}$ and $M_{\rm conf}$ of CRAFTS after two passes at DEC of $25.6 \degree$(in blue lines) and $-14 \degree$(in red lines). Solid lines represent $M_{\rm HI,lim}$ of CRAFTS, dashed lines represent $M_{\rm conf}$ including 2-Dimensional Correlation Function measured from $\alpha .40$ catalogue, which tells the excess probability(above random) of two galaxies being separated by a given distance in a rather small scale~\cite{Papastergis2013}. The solid-dashed lines represent the $M_{\rm conf}$ if we assume HI galaxies are uniformly distributed. The upper and bottom figure represents two different results by assuming $\Omega_{\rm HI}$ is constant and $\rho_{\rm HI} \, \propto \, (1+z)^{3}$ respectively. }
    \label{fig:Mconf}
\end{figure}

$M_{\rm conf}$ is closely related with the beamsize and the beamsize of FAST would vary with ZA and frequency. Fig.~\ref{fig:Mconf} shows the comparison of $M_{\rm conf}$ and $M_{\rm HI,lim}$ of two-pass CRAFTS at two typical DECs. It seems that confusion will not affect CRAFTS in detecting sources too much, because $M_{\rm conf}$ is smaller than $M_{\rm HI,lim}$ at two typical DECs. Perhaps it is due to the limit of integration time and that the beamsize of FAST at high redshift will be smaller than usual.

\section{Discussion and Conclusion}\label{sec:7}

In this paper, we quantify the science prospect of extragalactic HI detections based on a planned large-scale survey, namely, the Commensal Radio Astronomy FasT Survey (CRAFTS). We use FAST commissioning data to estimate parameters including the beamsize, the gain and the system temperature. We make a mock catalogue based on the HIMF derived from ALFALFA by assuming the HIMF doesn't evolve with redshift, and HI galaxies are distributed uniformly in the nearby universe. We also study the potential impact of confusion to CRAFTS survey using model in ~\cite{Jones2016a}. We summarized the expected results from CRAFTS as the following:

\begin{enumerate}

\item FAST plans to complete a blind drift scan search of HI galaxies as part of CRAFTS, which covers DEC between $-14 \degree$ and $66 \degree$, nearly $20000\, \rm deg^{2}$ of sky in two passes. The bandwidth of the receiver (FLAN) covers 1.05 GHz to 1.45 GHz, reaching a redshift of 0.35. The channel width is 7.6 kHz, which corresponds to a velocity resolution of approximately 1.6 $\rm km \, s^{-1}$ at HI rest frequency.

\item The main reflector of FAST can be fully iluminated when zenith angle is below $26.4 \degree$, with a beamsize of 2.95 arcmin at the HI rest frequency. At zenith angles above $26.4 \degree$, the mean reflector is partially illuminated and beamsize will increase linearly with zenith angle.

\item When zenith angle is below $26.4 \degree$, the gain of FAST is largely constant over the bandwidth of FLAN, which is close to 16.5 K/Jy. When zenith angle is above $26.4 \degree$, there will be a gain loss, at a zenith angle of $40 \degree$, the gain will decrease to approximately 11.0 K/Jy, with a loss of approximately $33\%$.

\item System temperature of FAST will increase with zenith angle ranging from approximately 18 K to 26 K.

\item The rms noise per 7.6 kHz channel of two-pass CRAFTS in most sky regions will be between 1.1 mJy and 1.5 mJy.

\item From our predictions, CRAFTS may detect over $6\, \times \, 10^{5}$ HI galaxies at a median redshift of 0.07 due to its wide coverage and outstanding sensitivity performance. With such huge number of samples, we can make unprecedented progress in studying HI survey products like the HI mass function (HIMF), the Correlation Function (CF) and the HI Velocity Width Function (WF) and deepening our understanding of HI distribution and its property in the universe.

\item  CRAFTS will not be confusion limited for detecting HI galaxy because of its limited integration time and unusual smaller beam size at high redshift.
\end{enumerate}

\vspace*{2mm} \Acknowledgements{\bahao This work is supported by National Key R$\&$D Program of China grant No.2017YFA0402600 and grant No.2016YFA0400702; the National Natural Science Foundation of China grant No.11690024, No.11725313, No.11590783 and No.11803051; the International Partnership Program of Chinese Academy of Sciences grant No.114A11KYSB20160008; CAS "Light of West China" Program; the Young Researcher Grant of National Astronomical Observatories, Chinese Academy of Sciences. The author would like to thank helpful discussion with Michael G. Jones.}

\end{multicols}


\begin{thebibliography}{99}

\bibitem{Roberts1975} Roberts M. S., in Sandage M., Kristan J., eds, \textit{Galaxies and the Universe}. (University of Chicago Press, Chicago, 1975), p. 309

\bibitem {Barnes2001} Barnes D. G. et al., MNRAS, 322, 486 (2001).

\bibitem{Meyer2004} Meyer M. J. et al., MNRAS, 350, 1195 (2004).



\bibitem{Giovanelli2005} Giovanelli R. et al., AJ, 130, 2598 (2005).

\bibitem{Haynes2011} Haynes, M. P., Giovanelli, R., Martin, A. M., et al. AJ, 142, 170 (2011).

\bibitem{Giovanelli2015} Giovanelli, R., $\&$ Haynes, M. P. A$\&$A Rv, 24, 1 (2015).

\bibitem{Haynes2018} Haynes, M. P., Giovanelli, R., Kent, B. R. et al., ApJ, 861, 49 (2018).

\bibitem{Briggs1990} Briggs F. H., AJ, 100, 999 (1990).

\bibitem{Schechter1976} Schechter P., ApJ, 203, 297 (1976).

\bibitem{Klypin1999} Klypin, A., Kratsov. A.V., Valenzuela, O. $\&$ Prada, F., ApJ 522, 82 (1999).

\bibitem{Moore1999} Moore, B., Ghigna, S., Governato, F. et al., ApJ 723, 1359 (1999).

\bibitem{Storrie2000} Storrie-Lombardi L. J., Wolfe A. M., ApJ, 543, 552 (2000).

\bibitem{Peroux2003}P$\rm \acute{e}$roux C., McMahon R. G., Storrie-Lombardi L. J., Irwin M. J., MNRAS, 346, 1103 (2003).

\bibitem{Zwaan2005} Zwaan M. A., Meyer M. J., Staveley-Smith L., Webster R. L., MNRAS, 359, L30 (2005).

\bibitem{Martin2010} Martin A. M., Papastergis E., Giovanelli R., Haynes M. P., Springob C. M., Stierwalt S., ApJ, 723, 1359 (2010).

\bibitem{Jones2018} Jones, M. G., Haynes, M. P., Giovanelli, R., $\&$ Moorman, C., MNRAS, 477, 2 (2018).

\bibitem{Obreschkow2009} Obreschkow D., Croton D., DeLucia G., Khochfar S., Rawlings S., ApJ, 698, 1467 (2009).

\bibitem{Duffy2012a} Duffy A. R., Kay S. T., Battye R. A., Booth C. M., Dalla Vecchia C., Schaye J., MNRAS, 420, 2799 (2012a).

\bibitem{Moorman2014}Moorman C. M., Vogeley M. S., Hoyle F., Pan D. C., Haynes M. P., Giovanelli R., MNRAS, 444, 3559 (2014).

\bibitem{Jones2016b}Jones M. G., Papastergis E., Haynes M. P., Giovanelli R., MNRAS, 457, 4393 (2016b).


\bibitem{Duffy2008} Duffy A. R., Battye R. A., Davies R. D., Moss A., Wilkinson P. N., MNRAS, 383, 150 (2008).

\bibitem{Duffy2012} Duffy A. R., Meyer M. J., Staveley-Smith L., Bernyk M., Croton D. J., Koribalski B. S., Gerstmann D., Westerlund S., MNRAS, 426, 3385 (2012).

\bibitem{Nan2006}Nan R., Sci. China G, 49, 129 (2006).

\bibitem{Nan2011}Nan R., et al., International Journal of Modern Physics D, 20, 989 (2011).

\bibitem{Li2018} D. Li, P. Wang, L. Qian, et al., IEEE. 19, 112 arXiv:1802.03709 (2019).

\bibitem{Jiang2019}P. Jiang et al. (FAST project), Sci. China-Phys. Mech. Astron, Commissioning progress of FAST, in press, (2019)

\bibitem{Jin2013}C. Jin, K, Zhu and J. Fan, et al., proceeding of the ISAP, Nanjing (2013).

\bibitem{Hogg1999} Hogg, D. W., ArXiv Astrophysics e-prints, astro-ph/9905116 (1999).

\bibitem{Meyer2017}Meyer, M., Robotham, A., Obreschkow, D., et al., PASA, 34, arXiv:1705.04210 (2017).

\bibitem{Jones2015} Jones M. G., Papastergis E., Haynes M. P., Giovanelli R., MNRAS, 449, 1856 (2015).

\bibitem{Jones2016a} Jones, M. G., Haynes, M. P., Giovanelli, R., $\&$ Papastergis, E., MNRAS, 455, 1574 (2016a).

\bibitem{Papastergis2013}Papastergis E., Giovanelli R., Haynes M. P., et al., ApJ, 776, 43 (2013).

\bibitem{LiJ2018}Li J., Wang Y.-G., Kong M.-Z., Wang J., Chen X., Guo R., RAA, 18, 003 (2018).




\end{thebibliography}
\end{document}